\documentclass[final,5p,times,twocolumn]{elsarticle}

\usepackage[T1]{fontenc}
\usepackage[utf8]{inputenc}

\usepackage{amsmath}
\usepackage{amsfonts}
\usepackage{amssymb}
\usepackage{amsxtra}
\usepackage{array}
\usepackage{color}
\usepackage{dcolumn}
\usepackage{enumerate}
\usepackage{graphicx}
\usepackage{hepunits}
\usepackage{xspace}
\usepackage{CJKutf8}

\definecolor{purple}{rgb}{0.5,0,0.5}
\definecolor{blue}{rgb}{0.0,0,0.9}
\definecolor{prdblue}{rgb}{0.133,0.118,0.498}
\usepackage[colorlinks=true, pdfstartview=FitV, linkcolor=prdblue, citecolor= prdblue, urlcolor=prdblue]{hyperref}

\usepackage[mathscr,scaled=1.15]{urwchancal}
\DeclareFontFamily{OT1}{pzc}{}
\DeclareFontShape{OT1}{pzc}{m}{it}%
{<-> s * [1.15] pzcmi7t}{}
\DeclareMathAlphabet{\mathpzc}{OT1}{pzc}{m}{it}

\newcommand{\Cfat}{{\mathbb C}}
\newcommand{\fatg}{{\rm{I}}\!\Gamma}

\biboptions{sort&compress}

\journal{Physics Letters B}

\begin{document}
\begin{CJK}{UTF8}{song}

\begin{frontmatter}

\title{$\,$\\[-7ex]\hspace*{\fill}{\normalsize{\sf\emph{Preprint no}. NJU-INP 068/22}}\\[1ex]
Schwinger mechanism for gluons from lattice QCD}

\author[Campinas]{A.\,C.~Aguilar}

\author[UPO]{F.~De Soto}

\author[UValencia]{M.\,N.~Ferreira}

\author[UValencia]{J.~Papavassiliou}

\author[UPO]{F.~Pinto-G\'omez}

\author[NJU,INP]{C.\,D.~Roberts}

\author[UHe]{J.~Rodr\'{\i}guez-Quintero}

%
\address[Campinas]{
University of Campinas - UNICAMP, Institute of Physics ``Gleb Wataghin'', 13083-859 Campinas, S\~ao Paulo, Brazil}

\address[UPO]{
Dpto. Sistemas F\'isicos, Qu\'imicos y Naturales, Univ.\ Pablo de Olavide, E-41013 Sevilla, Spain}

\address[UValencia]{
Department of Theoretical Physics and IFIC, University of Valencia and CSIC, E-46100, Valencia, Spain}

\address[NJU]{
School of Physics, Nanjing University, Nanjing, Jiangsu 210093, China}
\address[INP]{
Institute for Nonperturbative Physics, Nanjing University, Nanjing, Jiangsu 210093, China}

\address[UHe]{
Department of Integrated Sciences and Center for Advanced Studies in Physics, Mathematics and Computation, 
University of Huelva, E-21071 Huelva, Spain\\[1ex]
%
\href{mailto:cristina.aguilar@unicamp.br}{cristina.aguilar@unicamp.br} (A. C. Aguilar);
\href{mailto:Joannis.Papavassiliou@uv.es}{Joannis.Papavassiliou@uv.es} (J. Papavassiliou);
\href{mailto:cdroberts@nju.edu.cn}{cdroberts@nju.edu.cn} (C. D. Roberts);
\\[1ex]
Date: 2022 Nov 22\\[-6ex]
}

\begin{abstract}
Continuum and lattice analyses have revealed the existence of a mass-scale in the gluon two-point Schwinger function.  It has long been conjectured that this expresses the action of a Schwinger mechanism for gauge boson mass generation in quantum chromodynamics (QCD).  For such to be true, it is necessary and sufficient that a dynamically-generated, massless, colour-carrying, scalar gluon+gluon correlation emerge as a feature of the dressed three-gluon vertex.  Working with results on elementary Schwinger functions obtained via the numerical simulation of lattice-regularised QCD, we establish with an extremely high level of confidence that just such a feature appears; hence, confirm the conjectured origin of the gluon mass scale.
\end{abstract}

\begin{keyword}
continuum and lattice Schwinger function methods \sep
Dyson-Schwinger equations \sep
emergence of mass \sep
gluons \sep
quantum chromodynamics \sep
Schwinger mechanism of gauge boson mass generation
\end{keyword}

\end{frontmatter}
\end{CJK}

\section{Introduction}
%
The fact that Poincar\'e-invariant quantum gauge field theories can support the dynamical generation of a gauge-boson mass was first demonstrated sixty years ago \cite{Schwinger:1962tn, Schwinger:1962tp}.  In that case -- quantum electrodynamics with massless fermions in $D=2$ spacetime dimensions, QED$_2$ -- a mass scale is already present, \emph{viz}.\ the coupling, $e$, has mass-dimension one; and the ``photon'' acquires a mass $m_\gamma = e/\surd \pi$ as a consequence of the dynamical generation of a pole in the dimensionless vacuum polarisation scalar.  This is today referred to as the Schwinger mechanism (of gauge boson mass generation).  Referring to the usual Coulomb potential, which is linear in two dimensions, this gauge boson mass is often interpreted as an expression of very effective charge screening by a countable infinity of massless fermion$+$antifermion pairs that ensures the interactions between separated external charges are exponentially suppressed \cite[Sec.\,4.1]{Marciano:1977su}.  In fact, the fermions that appear in the defining Lagrangian of QED$_2$ vanish in solving the theory, being absorbed in the generation of the massive ``photon''.

Similar statements hold for QED$_3$: the charge-squared carries mass dimension one; but since that theory is not exactly solvable, only approximate numerical results are available.  Nevertheless, interactions between external charges are screened because a dynamically generated pole appears in the gauge boson vacuum polarisation  \cite{Appelquist:1988sr, Maris:1996zg, Bashir:2008fk, Bashir:2009fv, Braun:2014wja}.

Likewise for $D=3$ quantum chromodynamics (QCD$_3$).  As with all such $D=3$ models, QCD$_3$ is super-renormalisable, but it is \emph{a priori} plagued by infrared instabilities.  However, they may plausibly be cured by the dynamical generation of a gauge-boson mass via the Schwinger mechanism \cite[Ch.\,9]{Cornwall:2010upa}: this mass is also proportional to the gauge coupling squared.  Studies of gauge boson mass generation in QCD$_3$ have provided valuable insights, \emph{e.g}., Refs.\,\cite{Aguilar:2010zx, Cornwall:2015lna}.

$D=4$ quantum field theories, in general, and quantum chromodynamics (QCD), in particular, are different because, absent Higgs boson couplings, the classical Lagrangian is scale invariant: there are no intrinsic mass scales.  Nevertheless, the possibility that a Schwinger mechanism is active in QCD was conjectured forty years ago \cite{Cornwall:1981zr} and soon thereafter supported by a numerical simulation of lattice-regularised QCD (lQCD) \cite{Mandula:1987rh}.

The idea has since been refined \cite{Aguilar:2008xm, Boucaud:2011ug, Aguilar:2015bud, Binosi:2022djx, Papavassiliou:2022wrb}.  It is now known that a Schwinger mechanism is active if, and only if, a special type of longitudinally-coupled, negative-residue, simple-pole structure is dynamically generated in the three-gluon vertex.  Our focus, herein, is a demonstration, exploiting numerical results from lQCD, that just such a feature appears.  This being the case, then \cite{Gao:2017uox, Cui:2019dwv}: a gluon mass-scale is generated; the Landau pole is eliminated; and QCD is rendered infrared complete.  These are some of the consequences of emergent hadron mass (EHM) in the Standard Model \cite{Roberts:2020udq, Roberts:2020hiw, Roberts:2021xnz, Roberts:2021nhw, Binosi:2022djx, Papavassiliou:2022wrb, Ding:2022ows, Roberts:2022rxm}, the verification of which is being sought in an array of experimental programmes \cite{Aguilar:2019teb, Brodsky:2020vco, Carman:2020qmb, Chen:2020ijn, Anderle:2021wcy, Arrington:2021biu, Mokeev:2022xfo, Quintans:2022utc}.

\section{${\mathbb C}(r^2)$: keystone of Schwinger mechanism}
Following Refs.\,\cite{Schwinger:1962tn, Schwinger:1962tp, Cornwall:1981zr}, the natural place to begin a discussion of a dynamically generated gluon mass is the gluon two-point Schwinger function ($q^2 P_{\mu\nu}^q= \delta_{\mu\nu} q^2 - q_\mu q_\nu$):
\begin{equation}
\label{GluonSF}
\Delta_{\mu\nu}(q) = P_{\mu\nu}^q \frac{1}{q^2[ 1 +\Pi(q^2)] }
=:P_{\mu\nu}^q \, \Delta(q^2)\,.
\end{equation}
Landau gauge is used because it is a fixed point of the renormalisation group \cite[Ch.\,IV]{Pascual:1984zb} and readily implemented in lQCD \cite{Cucchieri:2009kk}.  Of course, gauge covariance of Schwinger functions ensures that all expressions of EHM in physical observables are independent of the gauge used for their elucidation.

\begin{figure}[!t]
\centerline{%
\includegraphics[clip, width=0.5\textwidth]{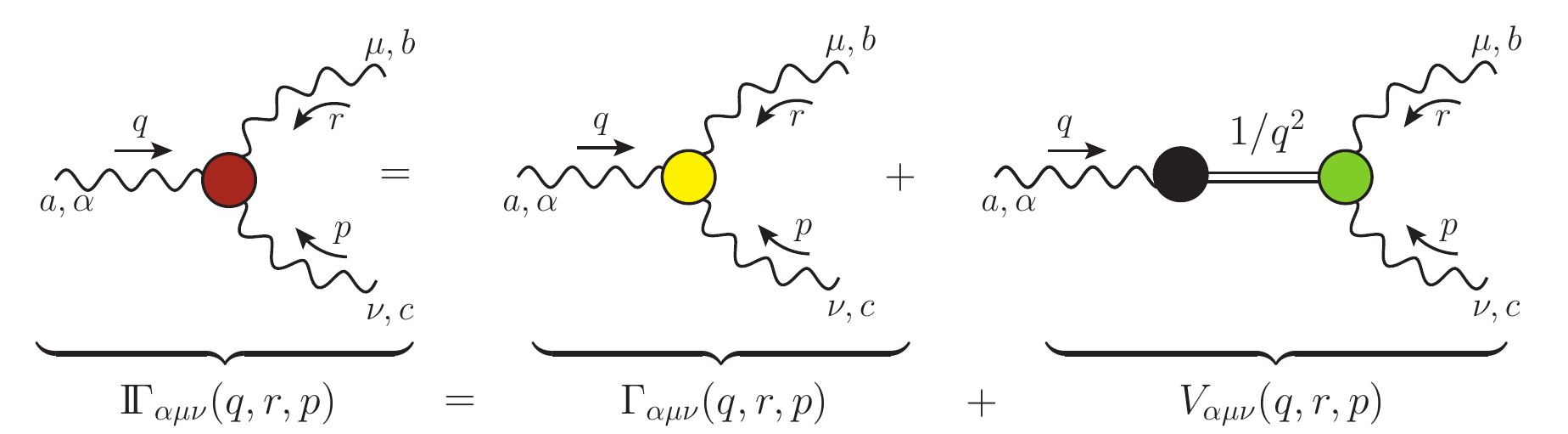}}
\caption{\label{F3gluon} Separation of the three-point gluon Schwinger function into a pole-free component plus a possibly nonzero part that exhibits a longitudinally-coupled simple-pole structure.  Undulating lines are gluons, filled circles are amputated vertices, and the double-line is the propagator of the putative massless colour-carrying scalar gluon+gluon correlation.}
\end{figure}

The dimensionless gluon self energy (vacuum polarisation), $\Pi(q^2)$ in Eq.\,\eqref{GluonSF}, may be obtained by solving the gluon gap equation, depicted, \emph{e.g}., in Ref.\,\cite[Fig.\,1]{Papavassiliou:2022wrb}.\footnote{Matter fields are omitted because, even when perturbatively massless, their impact on gauge boson mass generation is practically negligible \cite{Fischer:2003rp, Kamleh:2007ud, Aguilar:2012rz, Binosi:2016xxu, Cui:2019dwv}.}  Two of the five self-energy diagrams ($d_{1,4}$) involve the three-gluon vertex, which we write in the following form:
\begin{equation}
\fatg_{\alpha\mu\nu}(q,r,p) = \Gamma_{\alpha\mu\nu}(q,r,p) + V_{\alpha\mu\nu}(q,r,p) \,.
\label{fullgh}
\end{equation}
In our conventions, all momenta flow into the vertex, $r$ is the in-loop momentum, so $q+r+p=0$ --  see \,Fig.\,\ref{F3gluon}.

As highlighted by Fig.\,\ref{F3gluon}, Eq.\,\eqref{fullgh} separates this Schwinger function into two pieces, \emph{viz}.\ $\Gamma_{\alpha\mu\nu}$, which is the pole-free part that is usually considered, plus a (possibly) nonzero component that possesses a longitudinally-coupled simple-pole structure
\begin{equation}
V_{\alpha\mu\nu}(q,r,p) = \frac{q_\alpha}{q^2} \delta_{\mu\nu} C_1(q,r,p) + \ldots \,,
\label{eq:Vgen}
\end{equation}
where the ellipsis denotes analogous terms involving $r_\mu/r^2$, $p_\nu/p^2$, required by Bose symmetry, and other contributions that are subleading on $q^2\simeq 0$.  Notably, Bose symmetry of the three-gluon vertex also entails \cite{Aguilar:2021uwa} $C_1(q,r,p)+C_1(q,p,r)=0 \Rightarrow C_1(0,r,-r)=0$; hence,
\begin{subequations}
\label{eq:taylor_C}
\begin{align}
C_1(q,r,p) & \stackrel{q^2\simeq 0}{=} 2 q\cdot r \, {\mathbb C}(r^2) \, +  \, {\cal O}(q^2) \,, \\
{\mathbb C}(r^2) & := \left. \frac{\partial {C}_1(q,r,p)}{\partial p^2} \right|_{q = 0} \,. \label{KillerQueen}
\end{align}
\end{subequations}
The scalar function in Eq.\,\eqref{KillerQueen} is the keystone for a realisation of the Schwinger mechanism in QCD, playing a dual role: ${\mathbb C}(r^2)$ is both
\begin{enumerate}[(a)]
\item the amplitude associated with dynamical generation of a massless colour-carrying scalar gluon+gluon correlation;
\item and the \emph{displacement function} that quantifies modifications of the Ward identities satisfied by $\Gamma_{\alpha\mu\nu}(q,r,p)$, the pole-free part of the three-point gluon Schwinger function, in the presence of longitudinally-coupled massless poles. \label{Propb}
\end{enumerate}
We tacitly assume throughout that BRST symmetry \cite[Ch.\,II]{Pascual:1984zb} remains a feature of the solution of QCD so that all fully-dressed Schwinger functions satisfy their associated Slavnov-Taylor identities (STIs).

\begin{figure*}[!t]
\centerline{%
\includegraphics[clip, width=0.9\textwidth]{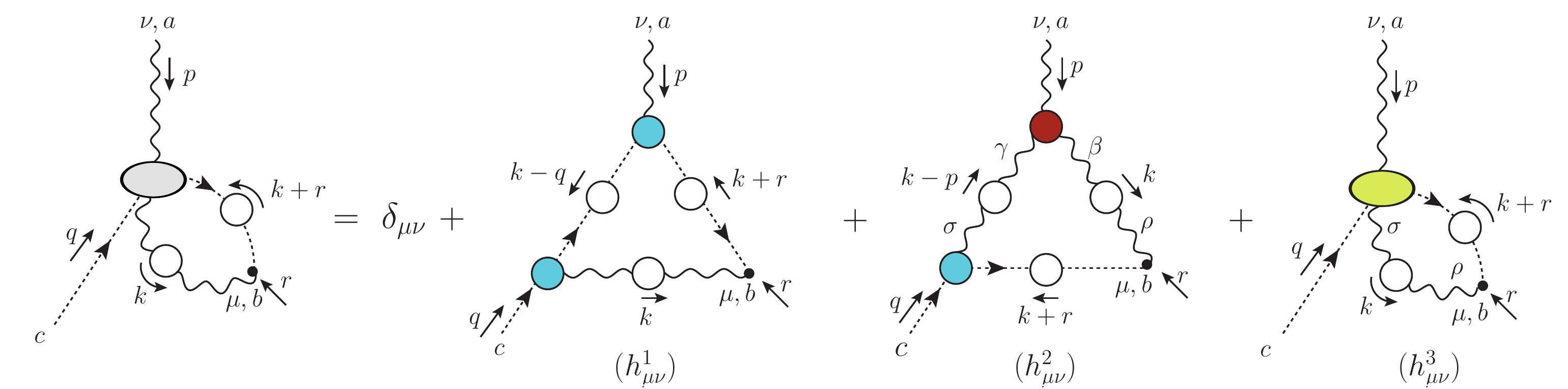}}
\caption{\label{H_DSE}
Dyson-Schwinger equation for the ghost-gluon scattering kernel, $H_{\mu\nu}(q,r,p)$: undulating lines are gluons, dotted lines are ghosts, and an open circle on such a line indicates a dressed-propagator; filled blue circles are amputated ghost-gluon vertices, the filled red circle is the analogous three-gluon vertex, and filled-ovals are scattering matrices.}
\end{figure*}

Pursuing property (\ref{Propb}) further, it was recently demonstrated \cite{Aguilar:2021uwa} that the displacement function can be expressed entirely in terms of elements that enter into the STI satisfied by $\fatg_{\alpha\mu\nu}$:
\begin{equation}
\Cfat(r^2) =
L_{sg}(r^2) - F(0)\left\{\frac{{\cal W}(r^2)}{r^2}\Delta^{-1}(r^2) + \widetilde{Z}_1 \,\frac{d \Delta^{-1}(r^2)}{dr^2 }\right\} \,.
\label{centeuc}
\end{equation}
Here:
the soft-gluon form factor, $L_{sg}$, expresses dynamics contained in a specific projection of the three-gluon vertex,
\begin{equation}
{P}_{\mu\mu^\prime}^r {P}_{\nu\nu^\prime}^{-r}\Gamma_{\alpha\mu^\prime\nu^\prime}(0,r,-r) = 2 L_{sg}(r^2) r_\alpha P_{\mu \nu}^r \,,
\label{TGamma}
\end{equation}
and may be extracted, \emph{e.g}., from lQCD results for the momentum space three-gluon Schwinger function $\langle {A}^a_\alpha(0) {A}^b_\mu(r) {A}^c_\nu(-r) \rangle$ \cite{Aguilar:2021lke, Aguilar:2021okw};
$\Delta(r^2)$ is defined in Eq.\,\eqref{GluonSF};
the ghost two-point function has been expressed as $D(q^2)=-F(q^2)/q^2$, so
$F(q^2)$ is the ghost dressing function, which satisfies $F(0)\in (0,\infty)$;
${\widetilde Z}_1$ is the ghost-gluon vertex renormalisation constant, whose Landau gauge properties are discussed elsewhere \cite{Taylor:1971ff}; and
\begin{equation}
\frac{{\cal W}(r^2)}{r^2} r_\rho \delta_{\mu\nu} =
\left. \frac{\partial H_{\mu\nu}(r,q,p)}{\partial q_\rho} \right|_{q=0} + \ldots\ \,,
\label{HKtens}
\end{equation}
where $H_{\mu\nu}$ is the ghost-gluon scattering kernel -- see Fig.\,\ref{H_DSE},
and the ellipsis indicates terms that do not contribute to Eq.\,\eqref{centeuc}.
%

A precise determination of ${\cal W}(r^2)$ from available lQCD results is a primary goal of our study because this enables calculation of the displacement function.

Three of the four functions appearing in Eq.\,\eqref{centeuc} are known with good precision from contemporary analyses of lQCD results \cite{Boucaud:2018xup}, \emph{viz}.\ $L_{sg}(r^2)$, $F(r^2)$, $\Delta(r^2)$; so that reliable fits and associated uncertainties are available -- see Refs.\,\cite[Fig.\,5]{Aguilar:2021lke}, \cite[Figs.\,4, 5]{Aguilar:2021okw}.  This is crucial because modern lQCD results for the gluon two-point function, $\Delta(k^2)$, also reveal the presence of a gluon mass scale \cite{Williams:2015cvx, Cyrol:2016tym, Binosi:2019ecz, Fischer:2020xnb, Falcao:2020vyr, Boito:2022rad}, but the lattice regularised theory is agnostic about its origin.  Thus, if one can use lQCD results alone to determine $\Cfat(r^2)<0$, then all practitioner-dependent bias is eliminated and the source of the gluon mass is unambiguously identified as the Schwinger mechanism.

The remaining quantity in Eq.\,\eqref{centeuc}, ${\cal W}(r^2)$, cannot be obtained directly from lQCD results.  Therefore, Ref.\,\cite{Aguilar:2021uwa} worked with a combination of lQCD output and an STI-inspired model for one primary element in the analysis to arrive at a lQCD-constrained form for ${\cal W}(r^2)$.  Using that to complete Eq.\,\eqref{centeuc}, a $\Cfat(r^2) < 0$ result was obtained whose form is in fair agreement with a solution of the related Bethe-Salpeter equation.  Herein, we go further by exploiting new lQCD results  \cite{Pinto-Gomez:2022brg} that can indirectly be used to determine ${\cal W}(r^2)$.

\section{Dyson-Schwinger equation for ${\cal W}(r^2)$}
Working from Eq.\,\eqref{HKtens}, one can derive an expression for ${\cal W}(r^2)$ in terms of the solution of the Dyson-Schwinger equation (DSE) for the ghost-gluon scattering kernel, drawn in Fig.\,\ref{H_DSE}.
Owing to transversality of the gluon two-point function, the ghost momentum, $q$, factors out of its radiative correction, enabling one to write \cite{Taylor:1971ff}
\begin{equation}
H_{\mu\nu}(q,r,p) = \tilde Z_1 \delta_{\mu\nu} + q_\rho K_{\mu\nu\rho}(r,q,p)\,;
\end{equation}
hence, using Eq.\,\eqref{HKtens},
\begin{equation}
{\cal W}(r^2) = \tfrac{1}{3} r_\rho P_{\mu\nu}^r K_{\mu\nu\rho}(r,0,-r)\,.
\end{equation}
It is worth recalling that $K_{\mu\nu\rho}(r,0,-r)$ is ultraviolet finite and its finite renormalisation is detailed in Ref.\,\cite{Aguilar:2020yni}.  Now, ${\cal W}(r^2)$ is readily obtained from the DSE solution.

There are three terms on the right-hand side of Fig.\,\ref{H_DSE}.  The third, $h_{\mu\nu}^3$, has been shown to contribute less than 2\% to the DSE's solution \cite{Huber:2017txg}.  Hence, we neglect it hereafter, writing
\begin{equation}
\label{WW1W2}
{\cal W}(r^2) = {\cal W}_1(r^2) + {\cal W}_2(r^2) \,,
\end{equation}
where ${\cal W}_{1,2}(r^2)$ are the contributions from $h^{1,2}_{\mu\nu}$, respectively.

The ghost-gluon vertex in Fig.\,\ref{H_DSE} is related thus to the ghost-gluon scattering kernel:
\begin{equation}
\label{EqGhostGlueV}
\Gamma_\nu(r,q,p) = r_\nu B_1(r,q,p) + p_\nu B_2(r,q,p) = r^\mu H_{\mu\nu}(r,q,p)\,,
\end{equation}
where, at tree-level, $B_1(r,q,p)=1$, $B_2(r,q,p)=0$ \cite{Aguilar:2018csq}.  Using Eq.\,\eqref{EqGhostGlueV} in Fig.\,\eqref{H_DSE}, one obtains
{\allowdisplaybreaks
\begin{subequations}
\label{EqW12}
\begin{align}
{\cal W}_1(r^2)= \tfrac{1}{2}&g^2\tilde Z_1 \int\!\tfrac{d^4k}{(2\pi)^4}\, \Delta(k^2) D(k^2) D(t^2)(r\cdot k ) \nonumber \\
& \times   B_1( t, - k , -r ) B_1(k,0,-k)  \left[ 1 - \frac{ (r \cdot k)^2 }{ r^2 k^2 } \right] ,\\
{\cal W}_2(r^2) = -\tfrac{1}{2}&g^2\tilde Z_1\int\!\tfrac{d^4k}{(2\pi)^4}\, \Delta(k^2) \Delta(t^2) D(t^2) \nonumber \\
&  \times B_1(t,0,-t) {\cal I}_{\cal W}(t^2,k^2,t^2)\,, \label{CalIW}
\end{align}
\end{subequations}
where $g$ is the strong coupling constant and $t=k+r$.
}

The hitherto undetermined elements in Eq.\,\eqref{CalIW} are
(\emph{i}) the contribution to ${\cal W}(r^2)$ from the three-point gluon Schwinger function, which appears as $\Gamma_{\nu \beta \gamma}^{a b c} (p , -k , p-k )$ in the $h_{\mu\nu}^2$ term of Fig.\,\ref{H_DSE},
and
(\emph{ii}) the ghost-gluon vertex function $B_1$.
Regarding (\emph{i}), with arguments and indices translated into the elementary form of the defining equation, Fig.\,\ref{F3gluon}, and implementing $q\to 0 \Rightarrow p\to -r $, then one is working with
\begin{equation}
{\cal I}_{\cal W}(r^2,k^2,t^2) = \tfrac{1}{2} (k-r)_\gamma \bar\Gamma_{\alpha\alpha\gamma}(-r,-k,k+r)\,,
\label{eq:IWdef}
\end{equation}
where \cite{Eichmann:2014xya, Aguilar:2019uob, Aguilar:2019kxz}
\begin{equation}
\label{T3gluon}
\bar\Gamma_{\alpha\mu\nu}(q,r,p) =
P_{\alpha \alpha^\prime}^q P_{\mu \mu^\prime}^r P_{\nu \nu^\prime}^p
\fatg_{\alpha^\prime\mu^\prime\nu^\prime}(q,r,p)\,.
\end{equation}
The remaining element, $B_1$, is discussed in Ref.\,\cite{Aguilar:2021okw}.  It may be obtained by solving a coupled pair of DSEs: that for the ghost-gluon vertex and the DSE for the ghost two-point function.  The kernels of these equations involve two functions known already from lQCD analyses -- $\Delta$, $D$; and $\bar\Gamma_{\alpha\mu\nu}$ in Eq.\,\eqref{T3gluon}, which must be determined.  Solving this pair of DSEs returns $\tilde Z_1$, $B_1$.

It is worth stressing that only the transverse projection of the three-gluon Schwinger function plays a role in determining ${\cal W}$.
For future reference, we record the tree level result:
\begin{align}
{\cal I}_{\cal W}^0(q^2, & r^2,p^2) =
\frac{1}{2p^2q^2r^2}\left[ 4 q^2 r^2 - \left(p^2 - q^2 - r^2 \right)^2\right] \nonumber \\
    &\times \left[ 3q^2r^2 - \frac{1}{4} \left(r^2-q^2-p^2\right)\left(q^2-r^2-p^2\right)\right] \;.
     \label{eq:IW0}
\end{align}

\section{Completing the kernel of the DSE for ${\cal W}(r^2)$}
We follow two paths to determining ${\cal I}_{\cal W}$.  The first -- M1 -- relies entirely on lQCD.  Specifically, the transverse projection of the three-gluon Schwinger function is directly accessible:
\begin{align}
g \bar\Gamma_{\alpha\mu\nu}(q,r,p)
= \frac{{\mathpzc G}_{\alpha\mu\nu}(q,r,p)}{\Delta(q^2)\Delta(r^2)\Delta(p^2)}\,,
\end{align}
where all elements in the ratio are obtained from simulations of the momentum space Schwinger functions
\begin{subequations}
\label{eq:Green2g}
\begin{align}
{\mathpzc G}_{\alpha\mu\nu}(q,r,p) & = \tfrac{1}{24} f^{abc} \langle A_\alpha^a(q) A_\mu^b(r) A_\nu^c(p) \rangle\,, \\
\Delta(q^2) & = \tfrac{1}{24} \delta^{ab} P_{\mu\nu}^q \langle A_\alpha^a(q) A_\mu^b(-q) \rangle\,.
\end{align}
\end{subequations}
A result for ${\cal I}_{\cal W}(r^2,k^2,t^2) $ follows immediately upon evaluating Eq.\,\eqref{eq:IWdef} using lQCD estimates of $\bar\Gamma_{\alpha\mu\nu}$.  We employ the lattice points obtained in Ref.\,\cite{Pinto-Gomez:2022brg}:
four sets of $2000$ configurations
generated on lattices with size $(L/a)^4= 32^4$
at bare couplings $\beta = 5.6$, $5.8$, $6.0$, $6.2$
so that $a=0.236, 0.144, 0.096, 0.070$, respectively, using the scale setting procedure in Ref.\,\cite{Boucaud:2018xup}.

As noted above, we adopt an asymmetric momentum subtraction renormalisation scheme \cite{Aguilar:2021lke, Aguilar:2021okw}, with renormalisation point $\zeta=4.3\,$GeV.  Regarding lQCD results for ${\cal I}_{\cal W}$, this proceeds by noting that
\begin{equation}
{\cal I}_{\cal W}(q^2,q^2,0) = - 6 q^2 L_{sg}(q^2)\,,
\label{IWexactL}
\end{equation}
having used Eq.\,\eqref{TGamma}; hence, both functions renormalise in the same way, \emph{i.e}.,
\begin{equation}
{\cal I}_{\cal W}^{\rm ren}(q^2,r^2,p^2) =  Z_3 {\cal I}_{\cal W}(q^2,r^2,p^2)\,,
\label{eq:renormalization}
\end{equation}
where $Z_3 L_{sg}(\zeta^2) = 1$.  Moreover, the renormalised gluon and ghost two-point functions are defined such that $\Delta^{-1}(\zeta^2)=\zeta^2$ and $F(\zeta^2) = 1$.  The latter entails $F(0) = 2.88$ \cite{Aguilar:2021lke}.

\begin{figure}[!t]
\vspace*{-2ex}

\leftline{\hspace*{0.5em}{\large{\textsf{A}}}}

\centerline{\includegraphics[clip, width=0.46\textwidth]{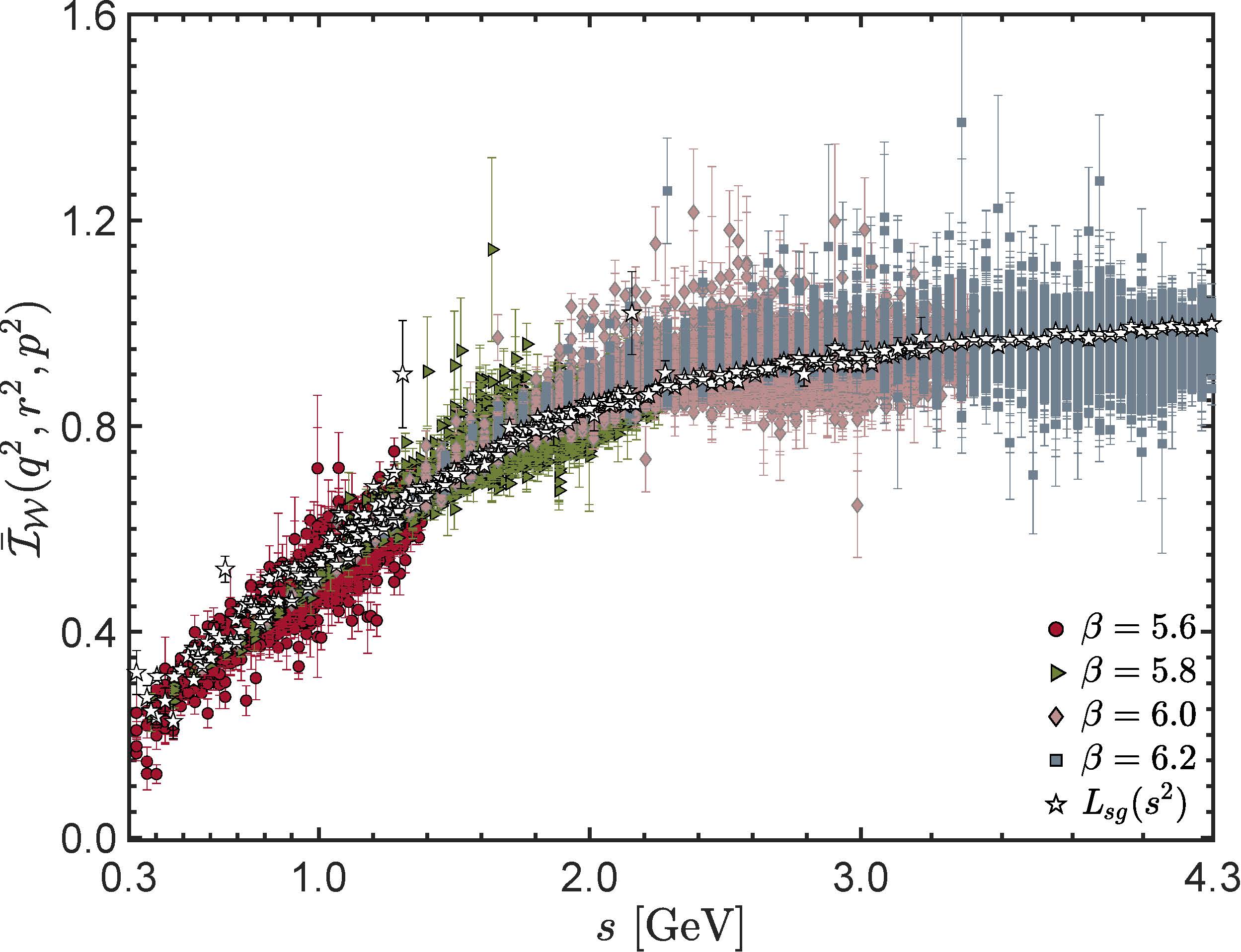}}

\leftline{\hspace*{0.5em}{\large{\textsf{B}}}}
\vspace*{-0ex}
\centerline{\includegraphics[clip, width=0.48\textwidth]{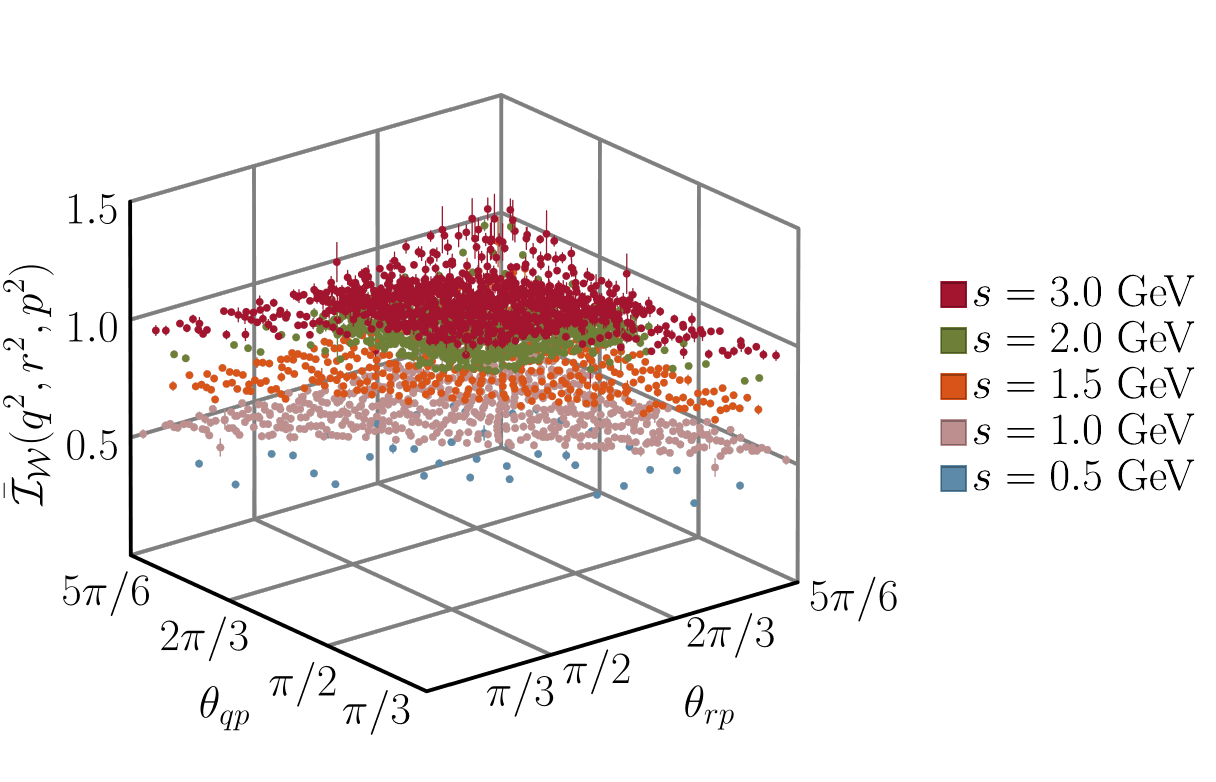}}
\caption{\label{fig:IWbar}
\emph{Upper panel}\,--\,{\sf A}.
Estimates for the ratio $\bar {\cal I}_{\cal W}$ in Eq.\,\eqref{eq:IWbar}, plotted as a function of the planar variable $s^2=(q^2+r^2+p^2)/2$ and obtained from lQCD results on all available configurations satisfying the kinematic constraints
$\theta_{qr}, \theta_{rp}, \theta_{pq} \leq 5\pi/6$, where $q\cdot r =: \sqrt{q^2 r^2} \cos  \theta_{qr}$, etc.,
and
$a(\beta) q, a(\beta) r, a(\beta) p \leq \pi/2$.
The open five-pointed stars are lQCD results for $L_{sg}(s^2)$ in Eq.\,\eqref{TGamma}.
\emph{Lower panel}\,--\,{\sf B}.
$\bar {\cal I}_{\cal W}$ plotted as a function of $(\theta_{rp},\theta_{qp})$ at
$s=3.0\,$GeV$^2$ -- red points, top plateau;
$s=2.0\,$GeV$^2$ -- olive points, next to top;
$s=1.5\,$GeV$^2$ -- orange points, middle;
$s=1.0\,$GeV$^2$ -- pink points, next to bottom;
$s=0.5\,$GeV$^2$ -- blue points, bottom plateau.
}
\end{figure}

We plot our lattice result for ${\cal I}_{\cal W}^{\rm ren}(q^2,r^2,p^2)$ in Fig.\,\ref{fig:IWbar}A, depicted via the ratio
\begin{equation}
\bar {\cal I}_{\cal W} (q^2,r^2,p^2) = {\cal I}_{\cal W}^{\rm ren} (q^2,r^2,p^2)/{\cal I}_{\cal W}^0 (q^2,r^2,p^2)\,.
\label{eq:IWbar}
\end{equation}

The ratio in Fig.\,\ref{fig:IWbar}A is plotted as a function of the symmetric, plateau variable $s^2=(q^2+r^2+p^2)/2$.  Evidently and importantly, as observed elsewhere \cite{Pinto-Gomez:2022brg}:
\begin{equation}
\bar {\cal I}_{\cal W} (q^2,r^2,p^2) = L_{sg}(s^2)\,,
\label{ApproxBarIW}
\end{equation}
within statistical precision.  This fact is emphasised by Fig.\,\ref{fig:IWbar}B, which depicts $\bar {\cal I}_{\cal W} (q^2,r^2,p^2)$ at fixed values of $s^2$ as a function of the direction cosines $(\theta_{rp},\theta_{qp})$: there is no statistically significant dependence on the angles.  Hence, one may interpret Eq.\,\eqref{ApproxBarIW} as delivering a sound approximation for ${\cal I}_{\cal W}^{\rm ren}(q^2,r^2,p^2)$; namely, the contribution to ${\cal W}$ from the three-gluon Schwinger function is reliably given by a product of the Bose-symmetric function $L_{sg}(s^2)$ with the tree-level result, Eq.\,\eqref{eq:IW0}, in which all angular dependence resides.  This approximation defines our second path -- M2 -- to determining ${\cal I}_{\cal W}$.

In order to employ M1, one requires a smooth interpolation of the lQCD results drawn in Fig.\,\ref{fig:IWbar}A.  Direct interpolation is unsuitable because those results are (\emph{a}) irregularly distributed on the momentum domain, having been obtained on four different lattices, and (\emph{b}) noisy.  Fitting is also unfavourable, given that one is working with a function of three variables, which makes it difficult to identify an optimal function form.

\begin{figure}[!t]
\centerline{%
\includegraphics[clip, width=0.4\textwidth]{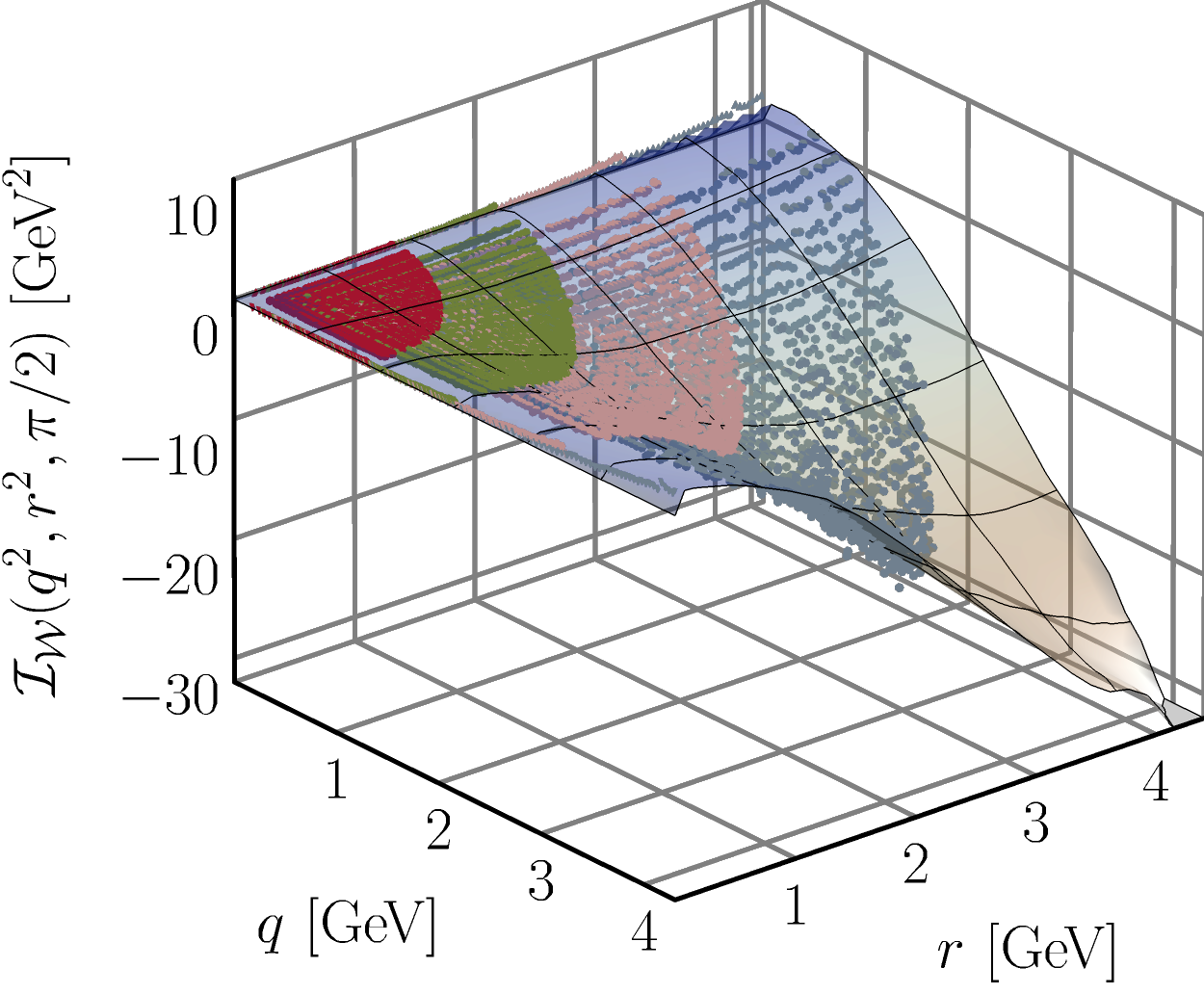}}
\caption{
Comparison between lattice results for ${\cal I}_{\cal W}(q^2,r^2,\theta_{qr})$ (dense points -- colour coded as in Fig.\,\ref{fig:IWbar}A) and our neural network predictor function (coloured surfaces), with $\theta_{qr} = \pi/2$ chosen for this illustration.
(Recall $q+r+p=0$, so the dependence on $p^2$ can be replaced by that on $\theta_{qr}$.)
\label{fig:IW_NN_vs_data}}
\end{figure}

To avoid these issues, we chose to employ a machine learning approach, training a neural network so as to obtain a continuous predictor function.  The algorithm is simple.
Beginning with the 335\,628 lQCD points in Fig.\,\ref{fig:IWbar}A, we randomly selected one-third as the training set.  Feeding that set to the Mathematica routine ``Predict'', with the ``NeuralNetwork'' option, we thereby obtained the desired predictor function.  The fidelity of the predictor function was gauged via comparisons with the remaining two-thirds of the points (223\,752): in no case did the predictor-function value differ by more than one standard-deviation from a test value.  The accuracy and smoothness of the predictor function is illustrated by Fig.\,\ref{fig:IW_NN_vs_data}.

\begin{figure}[!t]
\centerline{%
\includegraphics[clip, width=0.46\textwidth]{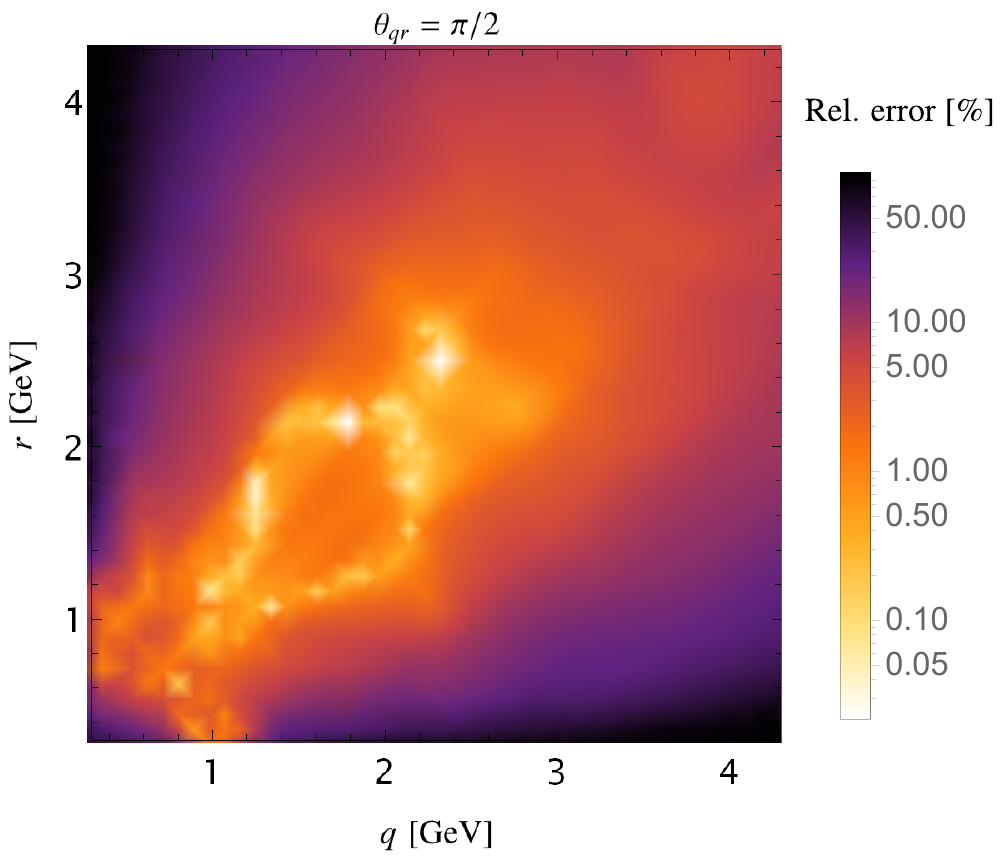}}
\caption{
Using Eq.\,\eqref{ApproxBarIW} as the definition of an approximation to $\bar{\cal I}_{\cal W}(q^2,r^2,\theta_{qr})$, this image represents its relative accuracy in comparison with the neural network predictor generated from lQCD output for $\bar{\cal I}_{\cal W}(q^2,r^2,\theta_{qr})$.
\label{fig:IW_error_density}}
\end{figure}

In Fig.\,\ref{fig:IW_error_density} we display the relative difference between the M2 approximation to ${\cal I}_{\cal W}(q^2,r^2,\theta_{qr})$ defined by Eq.\,\eqref{ApproxBarIW} and the M1 neural network predictor generated from lQCD output for this quantity.  On a broad neighbourhood of the diagonal ($q^2=r^2$), they agree within 1\%; and the agreement is better than 10\% on almost the entire domain.  Large relative discrepancies only exist far from $q^2=r^2$, where lattice results are sparse -- presenting a challenge for the neutral network approach -- and the value of ${\cal I}_{\cal W}$ is small, wherefore even a negligible absolute error may map into a large relative error owing to the small denominator.

Having once more confirmed the planar degeneracy property of the three-gluon Schwinger function \cite{Pinto-Gomez:2022brg}, we employ it in the calculation of $B_1$, thereby, herein, updating the analysis of Ref.\,\cite{Aguilar:2021okw}.  Specifically, akin to Eq.\,\eqref{eq:IWdef}, one writes
\begin{align}
\bar\Gamma_{\alpha\mu\nu}&(q,r,p)  = L_{sg}(s^2) P_{\alpha \alpha^\prime}^q P_{\mu \mu^\prime}^r P_{\nu \nu^\prime}^p \nonumber \\
& \times
[(q-r)_{\nu^\prime}\delta_{\alpha^\prime \mu^\prime }
+(r-p)_{\alpha^\prime} \delta_{\mu^\prime \nu^\prime} + (p-q)_\mu \delta_{\alpha^\prime \nu^\prime}  ]
\label{BarGamma0}
\end{align}
and completes the kernels of the $B_1$ DSEs using the lQCD results already discussed for each element involved.
Subsequently solving those DSEs, using the asymmetric momentum subtraction renormalisation scheme, with renormalisation scale $\zeta=4.3\,$GeV \cite{Aguilar:2021lke, Aguilar:2021okw}, we find ${\widetilde Z}_1 \approx 0.9333 (75)$ and a solution for $B_1$ that reproduces all available lQCD results \cite{Sternbeck:2006rd, Ilgenfritz:2006he, Aguilar:2021okw}.

\section{Strength of lattice signal for Schwinger mechanism}
There is no Schwinger mechanism in QCD if ${\mathbb C}(r^2)= {\mathbb C}_0(r^2)\equiv 0$; and one readily finds, using Eq.\,\eqref{centeuc}, that such an outcome requires
\begin{equation}
\label{EqW0}
{\cal W}(r^2) = {\cal W}_0(r^2) =
r^2 \Delta(r^2)\left[\frac{L_{sg}(r^2)}{F(0)} - \tilde Z_1 \frac{d\Delta^{-1}(r^2)}{dr^2}\right].
\end{equation}
Thus, the strength of any lattice signal for the Schwinger mechanism may be measured with respect to these null results.

\begin{figure}[t]
\centering
\includegraphics[width=0.46\textwidth]{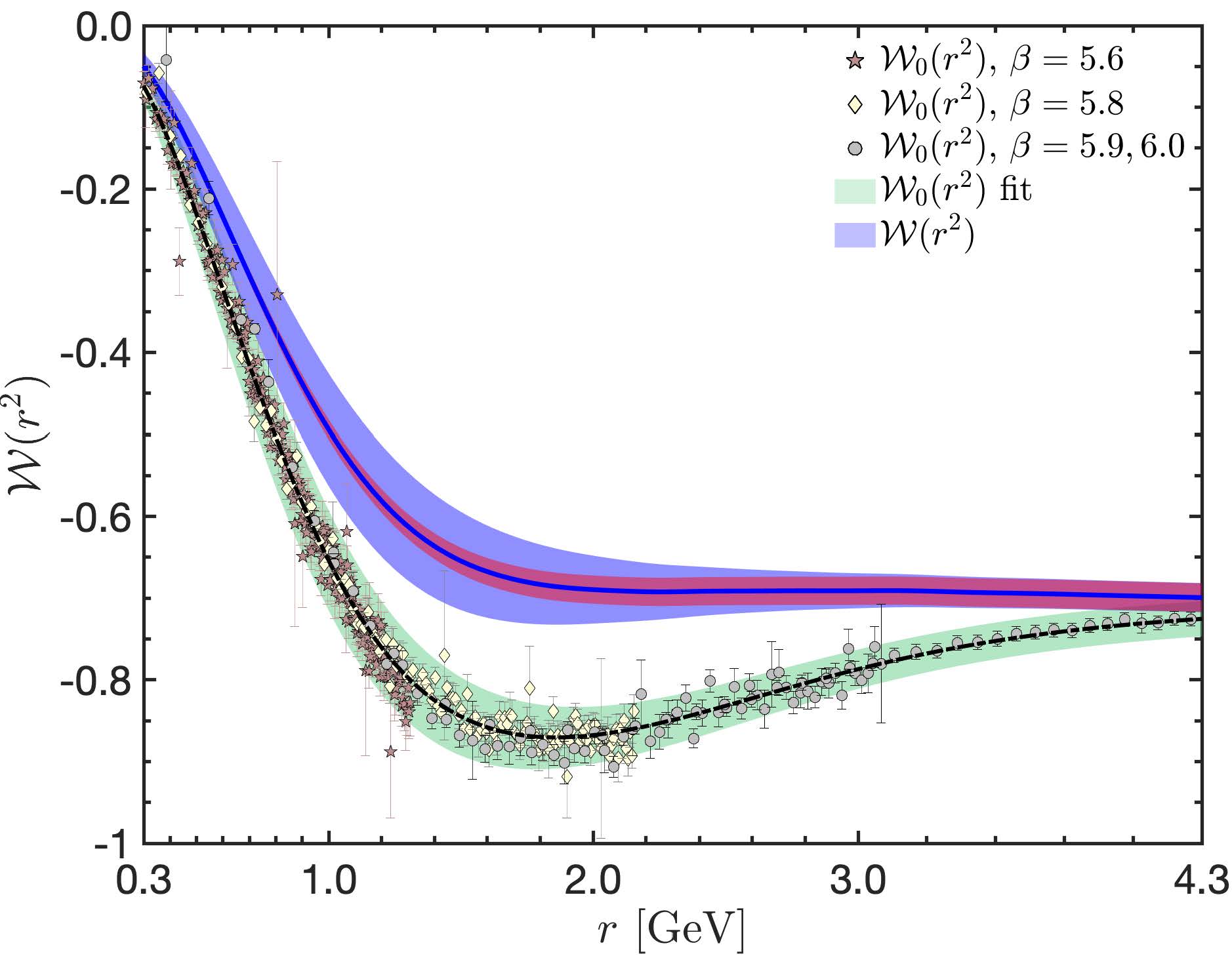}
\caption{\label{fig:W}
${\cal W}(r^2)$ calculated from Eqs.\,\eqref{WW1W2}, \eqref{EqW12} using lQCD-based inputs for every element, as described following Eq.\,\eqref{EqW0} -- solid blue curve.
An upper bound on the ${\cal W}(r^2)$-impact of systematic uncertainty in $L_{sg}(s^2)$ introduced by variations between the M1 and M2 approaches is indicated by the light-red band.
The total uncertainty, which combines that systematic uncertainty with the statistical error discussed in connection with Eq.\,\eqref{DeltaLsg}, is drawn as the blue band.
Null result (no Schwinger mechanism), ${\cal W}_0(r^2)$ in Eq.\,\eqref{EqW0} -- points computed using the lQCD values for $L_{sg}$ \cite{Aguilar:2021lke}, and dashed-black curve within the green uncertainty band, drawn using smooth fits to the lattice values.
}
\end{figure}

In order to determine ${\mathbb C}(r^2)$, one must first evaluate ${\cal W}(r^2)={\cal W}_1(r^2)+{\cal W}_2(r^2)$ using Eq.\,\eqref{EqW12}.
This requires knowledge of
$\Delta(k^2)$,
$D(k^2)$,
$L_{sg}(s^2)$,
and
${\cal I}(r^2,k^2,\theta_{rk})$,
$B_1(r^2,k^2,\theta_{rk})$
on $(k^2,\theta_{rk}) \in (0,\infty)\otimes (0,\pi)$.
For the first three, we use the fits to lQCD results described in Refs.\,\cite{Aguilar:2021okw, Aguilar:2021lke}, each of which was deliberately constructed so as reproduce the appropriate one-loop perturbative behaviour at ultraviolet momenta.

The last two elements -- ${\cal I}(r^2,k^2,\theta_{rk})$, $B_1(r^2,k^2,\theta_{rk})$ -- were determined above.  ${\cal I}(r^2,k^2,\theta_{rk})$ was calculated using two methods for the analysis of lQCD output, showing, too, that a reliable approximation to the lQCD results is provided by Eq.\,\eqref{ApproxBarIW}.  This justified our use of Eq.\,\eqref{BarGamma0} in computing $B_1$.
We employed these representations in our calculation of ${\cal W}(r^2)$, implemented using the fit to the lQCD result for $L_{sg}(s^2)$ discussed in Ref.\,\cite{Aguilar:2021lke}, which, as noted, ensures consistency with perturbative QCD.
Setting $g^2(\zeta=4.3\,{\rm GeV})/[4\pi] = 0.27$ \cite{Boucaud:2017obn}, our result for ${\cal W}(r^2)$ is displayed in Fig.\,\ref{fig:W}.

When using Eq.\,\eqref{ApproxBarIW}, one must propagate the lQCD statistical error on $L_{sg}(s^2)$ into an uncertainty on ${\cal W}(r^2)$.  Following Ref.\,\cite{Aguilar:2021uwa}, that may be achieved by introducing
\begin{equation}
\label{DeltaLsg}
L^\pm_{sg}(s^2) = L_{sg}(s^2) \pm \epsilon/[1+(r^2/\kappa^2)^2]\,,
\end{equation}
with $\epsilon=0.08$, $\kappa^2=5\,$GeV$^2$, and reevaluating ${\cal W}(r^2)$ using $L^\pm_{sg}(s^2)$.  Since it is improbable that lQCD uncertainties would lead to a uniform up/down shift in $L_{sg}(s^2)$, then this procedure establishes an upper bound on the associated uncertainty in ${\cal W}(r^2)$.
In this connection, it is important to recognise that all uncertainty in $B_1$ owing to Eq.\,\eqref{BarGamma0} is expressed in that associated with $\widetilde Z_1$.

\begin{figure}[t]
\centering
\includegraphics[width=0.46\textwidth]{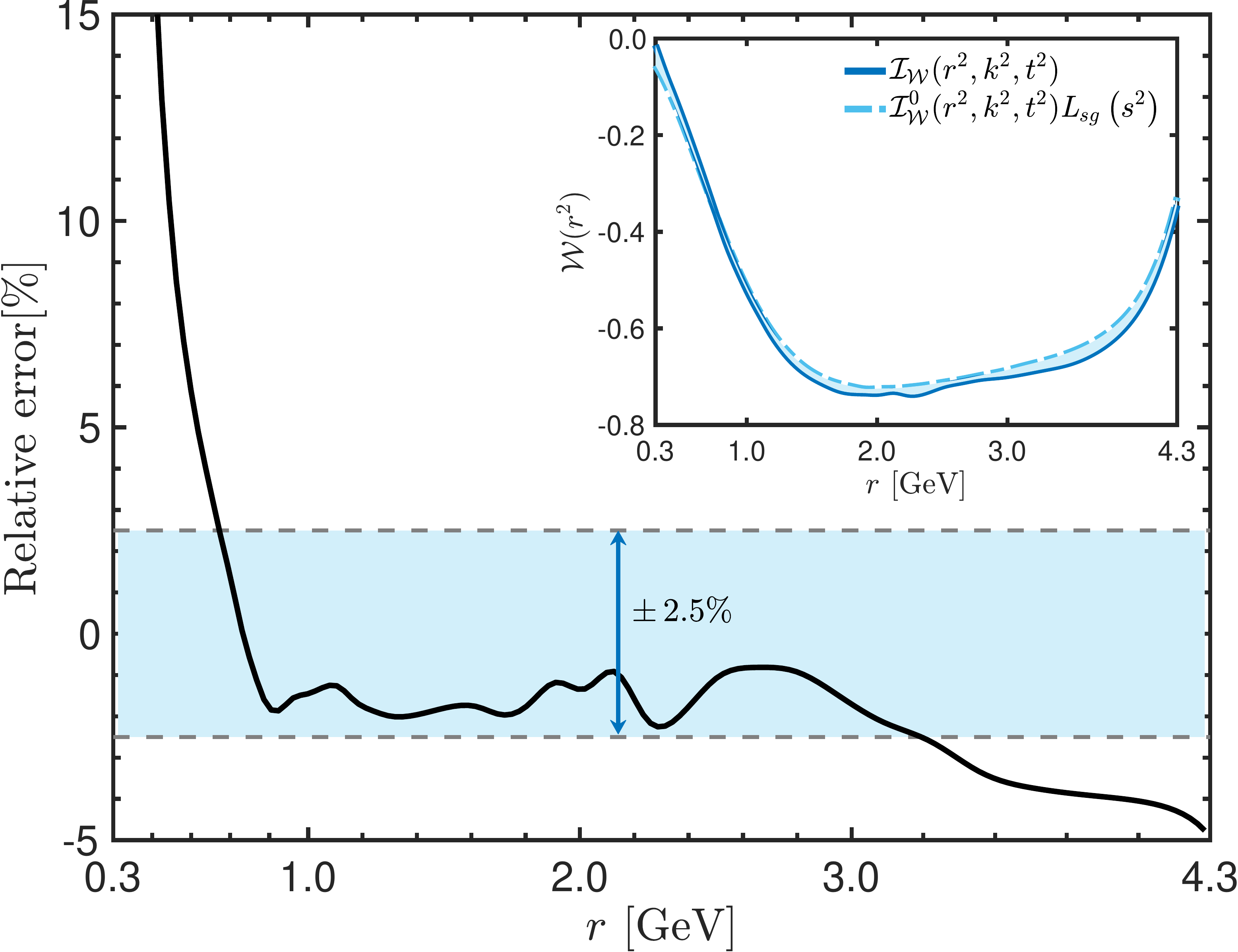}
\caption{\label{W_rest_error}
Relative deviation between contributions to ${\cal W}(r^2)$ from the integration subdomain $\surd k^2/{\rm GeV} \in [0.3,4.3]$ in Eqs.\,\eqref{WW1W2}, \eqref{EqW12} when ${\cal I}_{\cal W}$ is obtained using either Eq.\,\eqref{ApproxBarIW} or our neural network predictor function.
The blue band marks the $\pm 2.5$\% window.
The inset compares true values of the two results.
}
\end{figure}

In addition to the statistical error on $L_{sg}$, Eq.\,\eqref{ApproxBarIW} introduces a systematic uncertainty in the result for ${\cal W}(r^2)$ that can be quantified as follows.
(\emph{i}) In Eqs.\,\eqref{WW1W2}, \eqref{EqW12}, restrict the integration domain to $\surd k^2/{\rm GeV} \in [0.3,4.3]$, which is the subspace that contains almost all the available lQCD output for ${\cal I}_{\cal W}$.
(\emph{ii}) Integrating only over this subdomain, compare the values obtained for ${\cal W}(r^2)$ using Eq.\,\eqref{ApproxBarIW} (M2) with those obtained using the neural network predictor for ${\cal I}_{\cal W}$ (M1).
The resulting comparison is displayed in Fig.\,\ref{W_rest_error}.
Evidently, on almost the entire domain, the M2/M1 discrepancy propagated into ${\cal W}(r^2)$ is $\lesssim 2.5$\%.  It is significantly larger only at deep infrared momenta, which is that domain most sensitive to noise in ${\cal I}_{\cal W}$ introduced by finite lattice volume and expressed in the neutral network predictor.  Given its origin, the exaggerated error on this domain can be neglected.

It is here worth recalling Fig.\,\ref{fig:IW_error_density}, which shows domains whereupon the direct M2/M1 discrepancy is $\gg 2.5$\%.  Fig.\,\ref{W_rest_error} suggests strongly that those domains contribute little to the value of ${\cal W}(r^2)$.  Such may have been anticipated from Ref.\,\cite{Aguilar:2021uwa}, which showed that the integrand in Eq.\,\eqref{CalIW} is maximal on $t^2\simeq 0$, \emph{i.e}., $r^2=k^2$, $\theta_{rk}=\pi$.  On this domain, Eq.\,\eqref{ApproxBarIW} is exact -- see Eq.\,\eqref{IWexactL}.  The support of the integrand diminishes rapidly as $t^2$ increases, owing to its $\Delta(t^2) D(t^2)$ factor.  This ensures that those integration subdomains on which the evaluation of ${\cal I}_{\cal W}$ exhibits larger uncertainties -- deriving from the approximation and/or lQCD artefacts -- contribute little to the ${\cal W}(r^2)$ value.

The two sources of Eq.\,\eqref{ApproxBarIW}-related uncertainty in ${\cal W}$ that we have discussed are independent; hence, may be combined in quadrature.  This total uncertainty is drawn as the blue band in Fig.\,\ref{fig:W}: the systematic error (red band) dominates on $r\gtrsim 2\,$GeV, whereas the statistical uncertainty is most prominent on the complementary domain.  Notably, our lQCD calculation of ${\cal W}(r^2)$ yields a result very similar to that displayed in Ref.\,\cite[Fig.\,8]{Aguilar:2021uwa}, computed using an algebraic \emph{Ansatz} for the three-gluon Schwinger function \cite[Eq.\,(B10)]{Aguilar:2021uwa}.

\begin{figure}[t]
\centering
\includegraphics[width=0.46\textwidth]{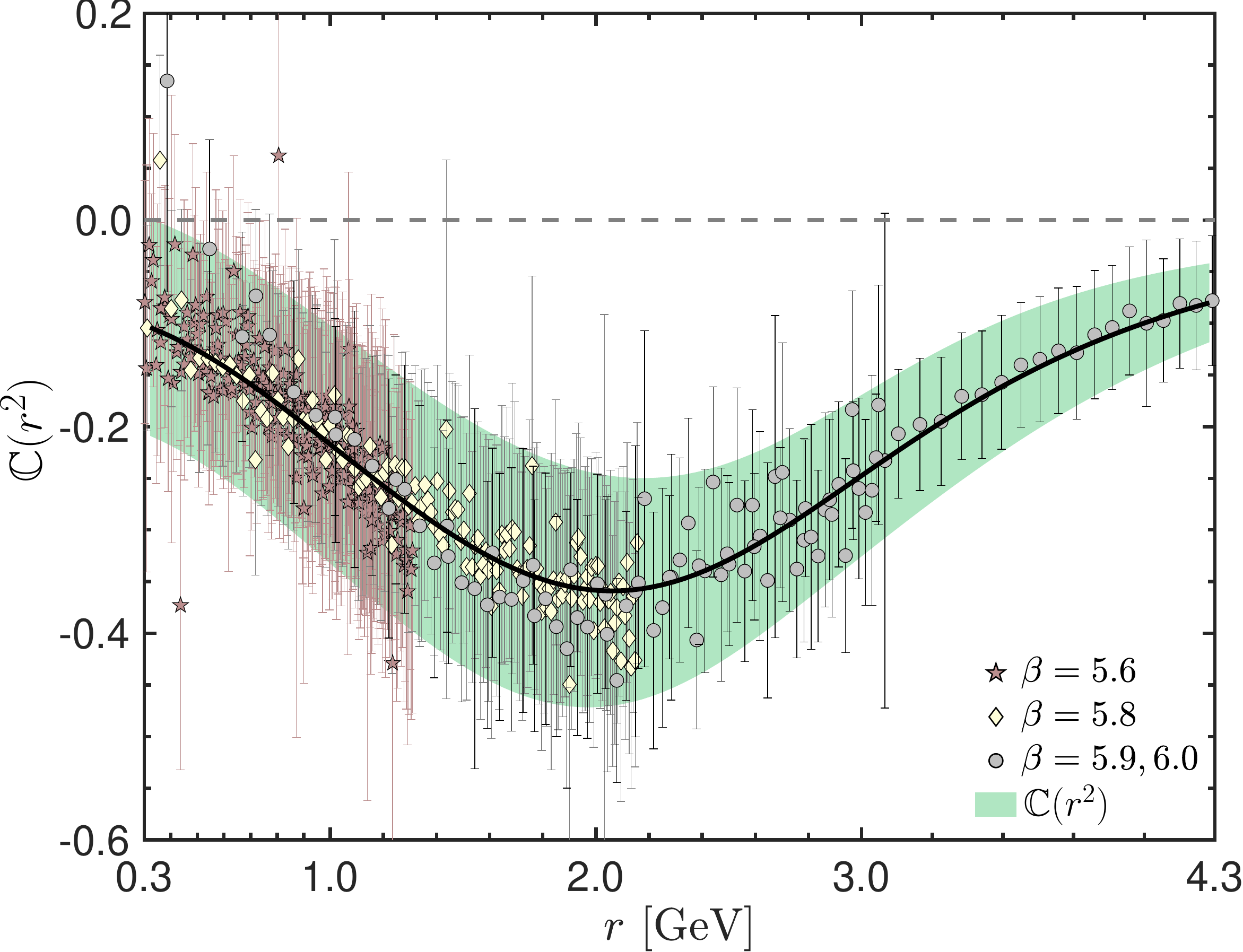}
\caption{\label{fig:Cfat}
Result for ${\mathbb C}(r^2)$.  Solid black curve -- obtained using central fit forms of lQCD results for the functions on the right-hand side of Eq.\,\eqref{centeuc}.  The bracketing soft-green band expresses the uncertainty in this result, discussed in connection with Eq.\,\eqref{CError}.
Dashed grey line -- null result (no Schwinger mechanism): ${\mathbb C}(r^2)={\mathbb C}_0(r^2)\equiv 0$.
The points represent ${\mathbb C}(r_i^2)$ obtained using explicit lQCD results \cite{Aguilar:2021lke} for the first term on the left-hand side of Eq.\,\eqref{centeuc}.
}
\end{figure}

At this point, we have in hand everything needed for evaluation of the Ward identity displacement function, $\mathbb C(r^2)$ in Eq.\,\eqref{centeuc}.  The result obtained, using central fit forms for the functions involved, is drawn as the solid black curve in Fig.\,\ref{fig:Cfat}.%
\footnote{As in perturbation theory, a careful analysis of the infrared behaviour of each element on the right-hand side of Eq.\,\eqref{centeuc} reveals that $|{\mathbb C}(0)|< \infty$, \emph{i.e}., all infrared divergences introduced by massless ghost loops cancel amongst themselves.}

The uncertainty in this result may be estimated by combining that in ${\cal W}(r^2)$ -- the blue band in Fig.\,\ref{fig:W} -- with the statistical errors on $L_{sg}(s^2)$.  (In comparison, published errors in lQCD results for the gluon two-point function are negligible.)
These two uncertainties are correlated because the calculation of ${\cal W}$ uses $L_{sg}$ as input; hence, they cannot be combined in quadrature.  In fact, there is a strong anticorrelation \cite{Aguilar:2021uwa}: increasing $L_{sg}(s^2)$ leads to a reduction in ${\cal W}(r^2)$ and vice versa.
A conservative bound on the uncertainty propagated into ${\mathbb C}(r^2)$ is therefore obtained by assuming maximal anticorrelation, in which case, including also the uncorrelated uncertainty on $\tilde Z_1$,
\begin{equation}
\label{CError}
\epsilon^2_{\mathbb C(r_i^2)} = \left[\rho_i + \tau_i \frac{F(0)}{r_i^2 \Delta(r_i^2)}\right]^2
+ \left[\delta_{\tilde Z_1} F(0) \frac{d \Delta^{-1}(r_i^2)}{dr_i^2}\right]^2
\,,
\end{equation}
where:
$\rho_i$ is the standard deviation in the lattice point for $L_{sg}(r_i^2)$ at the discrete lattice momentum values $r=r_i^2$ -- see Ref.\,\cite[Fig.\,5]{Aguilar:2021lke};
$\tau_i$ is the standard deviation in ${\cal W}(r_i^2)$, drawn from the blue envelope in Fig.\,\ref{fig:W};
and $\delta_{\tilde Z_1}=0.0075$.
The soft-green band bracketing the solid black curve in Fig.\,\ref{fig:Cfat} marks the extent of ${\mathbb C}(r_i^2) \pm \epsilon_{\mathbb C(r_i^2)}$, drawn using smooth interpolations of the upper and lower boundary points.

Figure~\ref{fig:Cfat} also displays results for ${\mathbb C}(r_i^2)$ obtained by using lQCD values for $L_{sg}(r_i^2)$ \cite{Aguilar:2021lke}, instead of the fit, to generate the first term on the left-hand side of Eq.\,\eqref{centeuc}.  The evident, significant agreement between the two methods boosts confidence in the central curve and uncertainty band.

It is now possible to quantify the significance of our ${\mathbb C}<0$ result as measured against the null hypothesis (no Schwinger mechanism): ${\mathbb C}(r^2)={\mathbb C}_0(r^2)\equiv 0$.
Consider, therefore
\begin{equation}
\label{NullCfat}
\chi^2 = \sum_{i=1}^{n_r} \frac{[{\mathbb C}(r_i^2) - {\mathbb C}_0(r_i^2)]^2}{\epsilon_{\mathbb C(r_i^2)}^2},
\end{equation}
where the sum runs over the $n_r=515$ values of $r_i \in [0.3,4.3]\,$GeV, for which the uncertainty in our result for ${\mathbb C}(r^2)$ is known, which evaluates to $\chi^2 = 2\,630$.  Consequently, the probability that our lQCD result for the displacement function is consistent with the null hypothesis is
\begin{subequations}
\label{PCfat0}
\begin{align}
{\rm P}_{{\mathbb C}_0} & = \int_{\chi^2=2\,630}^\infty \! dx\,\chi^2_{\rm PDF}(515,x) \\
& = \left. \frac{\Gamma(n/2,\chi^2/2)}{\Gamma(n/2)}\right|_{n=515}^{\chi^2=2\,630}= 7.3 \times 10^{-280}\,,
\end{align}
\end{subequations}
where we used $\chi^2_{\rm PDF}(2 n,2 x)=x^{n-1}\exp(-x)/(2^n \Gamma[n])$.
Even if the uncertainty on every value of ${\mathbb C}(r_i^2)$ were increased by 98\%, \emph{i.e}., practically doubled, the probability that our result could be consistent with the null hypothesis (no Schwinger mechanism) would still be less than $1/[1\,000\,000]$.

We note that the preceding analysis omits an assessment of the error introduced by neglecting $h_{\mu\nu}^3$ in the DSE for the ghost-gluon scattering kernel, Fig.\,\ref{H_DSE}.
Consider, therefore, that the null hypothesis is confirmed if, and only if, Eq.\,\eqref{EqW0} is satisfied, \emph{i.e}., ${\cal W}(r^2) = {\cal W}_0(r^2)$.
This ``null ${\cal W}''$ is represented by the points in Fig.\,\ref{fig:W}, which were computed using the lQCD values for $L_{sg}$ in Ref.\,\cite{Aguilar:2021lke}, and also the dashed-black curve within the green uncertainty band, drawn using smooth fits to the lattice values.
Using Eqs.\,\eqref{centeuc}, \eqref{EqW0} and carefully treating uncertainty correlations, it is readily established that the $\chi^2$ value for the ${\cal W}= {\cal W}_0$ hypothesis is also given by Eq.\,\eqref{NullCfat} and the associated realisation probability by Eq.\,\eqref{PCfat0}.
Now, since any contribution from $h_{\mu\nu}^3$ to the ghost-gluon vertex is less than 2\% \cite{Huber:2017txg}, then it cannot affect this probability, \emph{viz}.\ neglecting $h_{\mu\nu}^3$ has no measurable impact.

In closing this section it is worth stressing that should an alternative origin for the gluon mass scale be proposed, then, without artefice, it must simultaneously explain and reproduce the Ward identity displacement function in Fig.\,\ref{fig:Cfat}, which, as we have shown, is a feature of QCD.  Failing that, then the viability of the alternative may reasonably be challenged.

\section{Conclusion}
%
Working solely with lattice-QCD results for
(\emph{a})
the ghost two-point Schwinger function,
ghost-gluon vertex,
and gluon two-point function \cite{Aguilar:2021okw},
and (\emph{b}) the gluon three-point function \cite{Aguilar:2021lke, Pinto-Gomez:2022brg}, we calculated the Ward identity displacement function ${\mathbb C}$ [Fig.\,\ref{fig:Cfat}].
Were ${\mathbb C}\equiv 0$, then gluons could not acquire a mass via the Schwinger mechanism.  On the other hand, a ${\mathbb C}(r^2)< 0$ result signals that the gluon three-point function possesses a longitudinally-coupled, simple pole structure associated with a dynamically-generated, massless, colour-carrying, scalar gluon+gluon correlation; and this is necessary and sufficient to ensure that gluons acquire a (momentum-dependent) mass dynamically via the Schwinger mechanism.
Our analysis reveals that the ${\mathbb C}\equiv 0$ result is excluded with p-value $p=1-7.3 \times 10^{-280}$  [Eq.\,\eqref{PCfat0}], \emph{i.e}., with p-value unity by any reasonable assessment.
One may therefore conclude that a Schwinger mechanism is active in QCD, leading to the emergence of a gluon mass-scale through the agency of a dynamically-generated pole in the gluon three-point function.
A continuing effort is underway to expose measurable consequences of these phenomena.

\medskip
\noindent\textbf{Acknowledgments}.
We are grateful for constructive comments from D.~Binosi and Z.-F.~Cui.
Use of the computer clusters at the Univ.\ Pablo de Olavide computing centre, C3UPO, is gratefully acknowledged.
Work supported by:
National Natural Science Foundation of China (grant no.\,12135007);
National Council for Scientific and Technological Development (CNPq)  (grant  no.\, 307854/2019);
Spanish Ministry of Science and Innovation (MICINN) (grant nos.\ PID2019-107844GB-C22, PID2020-113334GBI00);
Generalitat Valenciana (grant no.\ Prometeo/2019/087);
and
Junta de Andaluc{\'{\i}}a (grant nos.\ P18-FR-5057, UHU-1264517).


\medskip
\noindent\textbf{Declaration of Competing Interest}.
The authors declare that they have no known competing financial interests or personal relationships that could have appeared to influence the work reported in this paper.



\end{document}